# Preliminary Goniopolarimetric Simulations for Space Weather Radio (SWARA) Polarimeter

Poorva Singh*(1), Mugundhan V.(1), and Anshu Kumari(2)
(1) Indian Institute of Technology, Kanpur, India, 208016, e-mail: poorvasi25@iitk.ac.in; mugundhanv@iitk.ac.in
(2) Physical Research Laboratory, Udaipur, India, 313001, e-mail: anshu@prl.res.in

## Abstract

Goniopolarimetry is a technique that measures the direction-of-arrival (DoA) using the polarization properties of the incident electromagnetic (EM) waves. This is done by using cross-correlations to estimate the projection of the polarization ellipse across multiple co-located orthogonal antennas. This study aims to incorporate the technique of Goniopolarimetry for our proposed polarimeter, Space Weather Radio (SWARA) Polarimeter, to provide early warnings for space weather or solar events. We have performed preliminary simulations for a tri-monopole antenna configuration. The behaviour of the tri-monopole antenna with varying signal-to-noise (SNR) ratios along the field-of-view (FOV) was investigated using simulations. The angular localization error (ALE) variations, along with SNRs, were simulated to quantify the sensitivity of the Goniopolarimetric algorithm to noise (Galactic, Receiver, and Antenna/System). The simulations were performed using MATLAB, taking the effective length of the monopoles to be 1 m. The present simulations establish a baseline for evaluating the goniopolarimetric performance of an ideal tri-monopole antenna, and the results obtained from these investigations will serve as the foundation for subsequent stages of the research. Future work will focus on determining the optimum effective length of the individual monopoles to maximize sensitivity and localization accuracy while ensuring efficient impedance matching with the receiver. Based on these findings, the design and optimization of a practical tri-dipole antenna system will be carried out to achieve broadband operation and improved electric vector field measurements. Furthermore, the optimized antenna will be evaluated under realistic plasma environments through electromagnetic and plasma simulations to investigate the influence of plasma parameters on antenna characteristics, received signal quality, and DoA estimations.

## 1. Introduction

Goniopolarimetry is a radio direction-finding technique. It utilizes the measurements of the electric field polarization of the incident electromagnetic (EM) waves, with the directional characteristics of multiple antennas to determine the Direction of Arrival (DoA). This technique exploits both the amplitude and polarization state of the received signal. Since the received signal polarization depends on the orientation of the incident wave relative to the antenna geometry, the DoA can be inferred without requiring large antenna arrays [1]. This makes goniopolarimetry particularly suitable for spaceborne and planetary radio observations. This technique enables accurate estimation of the incident wave vector using compact multi-component antenna systems such as crossed dipoles or tripole antennas [2].

The fundamental principle of goniopolarimetry is based on measurement of the three-dimensional electric field components of the EM wave. Then, by analyzing the covariance or spectral matrix of these measurements, the wave polarization parameters and propagation direction can be estimated [3]. This technique is especially advantageous at low radio frequencies, where constructing large interferometric baselines is impractical [4]. Apart from goniopolarimetry, vector antenna-based DoA estimation techniques, like Vector MUSIC, Vector ESPRIT, and Maximum Likelihood (ML) methods, exploit both the spatial and polarization information of the received EM field [5]. Unlike these techniques that typically require multiple vector sensors or arrays and involve higher computational complexity, goniopolarimetry can estimate the direction of arrival using a single vector antenna from the measured field vectors.

Goniopolarimetric systems commonly employ tripole or orthogonal dipole antennas that measure all three electric field components The measured voltages are processed to construct the covariance matrix, from which polarization characteristics such as ellipticity, degree of polarization, and wave normal direction are derived [3][6]. These parameters enable accurate determination of the incoming signal's DoA while remaining robust under varying signal polarization conditions.

Consequently, goniopolarimetry offers a compact and computationally efficient solution, making it particularly suitable for spaceborne applications where size, power, and processing resources are limited. It is a tool for radio astronomy, planetary exploration, ionospheric remote sensing, and space weather research, where precise localization of naturally occurring radio emissions is essential [7].

This study aims to incorporate the technique of Goniopolarimetry for our proposed polarimeter design. The Space Weather Radio (SWARA) Polarimeter is designed to provide early warnings for space weather or solar events such as Coronal Mass Ejections (CMEs), Solar Flares, and solar energetic particles. The payload is proposed to operate from 300 kHz to 30 MHz frequency range and will consist of deployable tri-axial antennas. The simulations generated in this study are used to establish a baseline for evaluating the goniopolarimetric performance of an ideal tri-monopole antenna.

## 2. Literature

### 2.1 Background Noise

The sensitivity of a radio receiving system is limited by the background noise present in the observing frequency band. At low radio frequencies, the dominant noise sources differ significantly. That makes accurate estimation of the background noise essential for evaluating the SNR. For frequencies below ~ 30 MHz, the dominant contribution to the system noise is from diffuse Galactic emission [8]. This emission originates from relativistic electrons spiralling in the Galactic magnetic field and forms a radio background that can be a dominant source of noise for space and ground-based radio telescopes. The galactic background exhibits a power-law dependence with frequency [9], and can be expressed as,

$$T_{gal} = 1.2 \times 10^7 \left(\frac{f}{1\,MHz}\right)^{-0.76} \tag{1}$$

$T_{gal}$ – Galactic background brightness temperature, K
f – Frequency, Hz

Along with the galactic background, Quasi-Thermal Noise (QTN) also contributes to the background noise. QTN is a natural EM noise generated by the random thermal motion of charged particles, primarily electrons, in a plasma. As electrons move randomly due to their thermal energy, they produce microscopic fluctuations in electric charge density and electric fields. These fluctuations induce voltages on the antenna, which are detected as noise [10]. It becomes a significant background noise source in hot space plasmas with electron temperatures typically in the range of $10^5$–$10^6$ K. For example, the solar wind sampled by Solar Orbiter, where it can dominate the receiver noise near the local plasma frequency [11].

QTN is strongest around the local electron plasma frequency ($f_p$), and its influence decreases rapidly at frequencies well above $f_p$. For most interplanetary missions, QTN is a dominant noise source only in the kHz range, typically below about 100 kHz, where the observation frequency is close to the local plasma frequency [12]. At MHz frequencies, QTN decreases significantly, and other noise sources, such as galactic background noise and receiver noise, generally dominate [10].

$$f_p = 8980\sqrt{n_e} \tag{2}$$

$f_p$ - Electron plasma frequency, Hz
$n_e$ – Electron density, $cm^{-3}$

Finally, the internal noise generated by the electronic components of the receiving system is also known as Receiver noise, which originates from thermal agitation and imperfections in the receiver circuitry, along with the random thermal motion of charge carriers in resistors, amplifiers, mixers, analog-to-digital converters, and other electronic circuits. It establishes the fundamental noise floor that limits the sensitivity and signal-to-noise ratio of the system. This noise exists even in the absence of any external radio signal [9].

In practice, the total system noise temperature is the sum of all significant noise contributions,

$$T_{bkg} = T_{gal} + T_{qtw} + T_r \tag{3}$$

$T_{bkg}$ – Background temperature, K
$T_{gal}$ – Galactic background brightness temperature, K
$T_{qtw}$ – Quasi-Thermal Noise temperature, K
$T_r$ – Receiver temperature, K

The total Background Noise power is therefore,

$$P_n = k_b T_{bkg}(BW) \tag{4}$$

$P_n$ – Background noise power, W
$k_b$ – Boltzmann constant ($1.380649 \times 10^{-23}$), $JK^{-1}$
$T_{bkg}$ – Background temperature, K
BW – Receiver bandwidth, Hz

## 2.2 Signal-to-Noise Ratio

The Signal-to-Noise Ratio (SNR) is a measure of how well the received radio signal stands out above the radio noise. It impacts the ability of an antenna to detect and localize an incoming radio signal under ideal space conditions [9]. The SNR is expressed as,

$$SNR = \frac{P}{P_n} \tag{5}$$

$$SNR_{dB} = 10 \log_{10}\left(\frac{P}{P_n}\right) \tag{6}$$

P – Received signal power, W
$P_n$ – Background noise power, W

## 2.3 Direction-of-Arrival

The Direction of Arrival (DoA) is the direction from which an EM wave reaches the receiving antenna [3]. For a plane electromagnetic wave propagating in free space, the direction of propagation is represented by the unit wave vector,

$$k = \begin{bmatrix} \sin\theta \cos\phi \\ \sin\theta \sin\phi \\ \cos\theta \end{bmatrix} \tag{7}$$

k – Direction of propagation of EM wave
θ – Elevation angle, degree/radian
ϕ – Azimuth angle, degree/radian

The incident electric field induces voltages on the three orthogonal monopoles, which are then combined to form the received signal/voltage vector [13],

$$V = \begin{bmatrix} V_x \\ V_y \\ V_z \end{bmatrix} \tag{8}$$

The above vector is used to calculate the covariance matrix of the received signals. The covariance matrix contains information about the amplitudes, relative phases, and polarization of the received wave. Here, $V^*$ is the Hermitian transpose of V, the voltage vector,

$$R = [V \times V^*] \tag{9}$$

Using goniopolarimetric inversion, the polarization parameters and the propagation vector are estimated. The complex electric field vector is first reconstructed from the measured antenna voltages using the calibrated antenna response matrix. The unit propagation vector is then obtained from the normalized imaginary part of the cross product of the electric field vector and its complex conjugate, using Equation 9 [13]. The solution corresponding to the minimum residual error is selected as the estimated Direction of Arrival (DoA).

The resulting propagation vector is expressed as [14],

$$\hat{k} = \frac{Img(|V \times V^*|)}{|Img(V \times V^*)|} \tag{10}$$

From the above equation, the estimated azimuth and elevation angles are obtained as,

$$\hat{\theta} = \cos^{-1} \hat{k}_z \tag{11}$$

$$\hat{\phi} = \tan^{-1} \frac{\hat{k}_y}{\hat{k}_z} \tag{12}$$

## 2.4 Angular Localization Error

The Angular Localization Error (ALE) is a quantitative measure of the accuracy of DoA estimation. It represents the angular separation between the true propagation direction of the incoming EM wave and the direction estimated by the goniopolarimetric algorithm [9]. ALE is one of the primary performance metrics for evaluating the localization capability of antenna systems. It can be calculated as the angle between the true propagation vector and the estimated propagation vector,

$$\omega = \cos^{-1}(\hat{k}_{true}.\hat{k}_{est.}) \tag{13}$$

ω – Angular Localization Error, degree

Alternatively, the ALE can also be estimated as [9],

$$\omega = \sqrt{(\hat{\theta} - \theta)^2 + \sin^2\theta\,(\hat{\phi} - \phi)^2} \tag{14}$$

It represents the angular difference between the true propagation direction of an incident EM wave and the direction estimated from the antenna measurements. A smaller ALE indicates that the estimated DoA is closer to the actual DoA, whereas a larger ALE signifies a less accurate localization.

## 3. Methodology

The simulations were carried out with some assumptions to evaluate the ideal performance of the tri-monopole antenna (as shown in Figure 1) for DoA estimation. The antenna consists of three mutually orthogonal monopoles aligned along the x-, y-, and z-axes [13], where all monopoles are identical with equal gain and identical radiation characteristics.

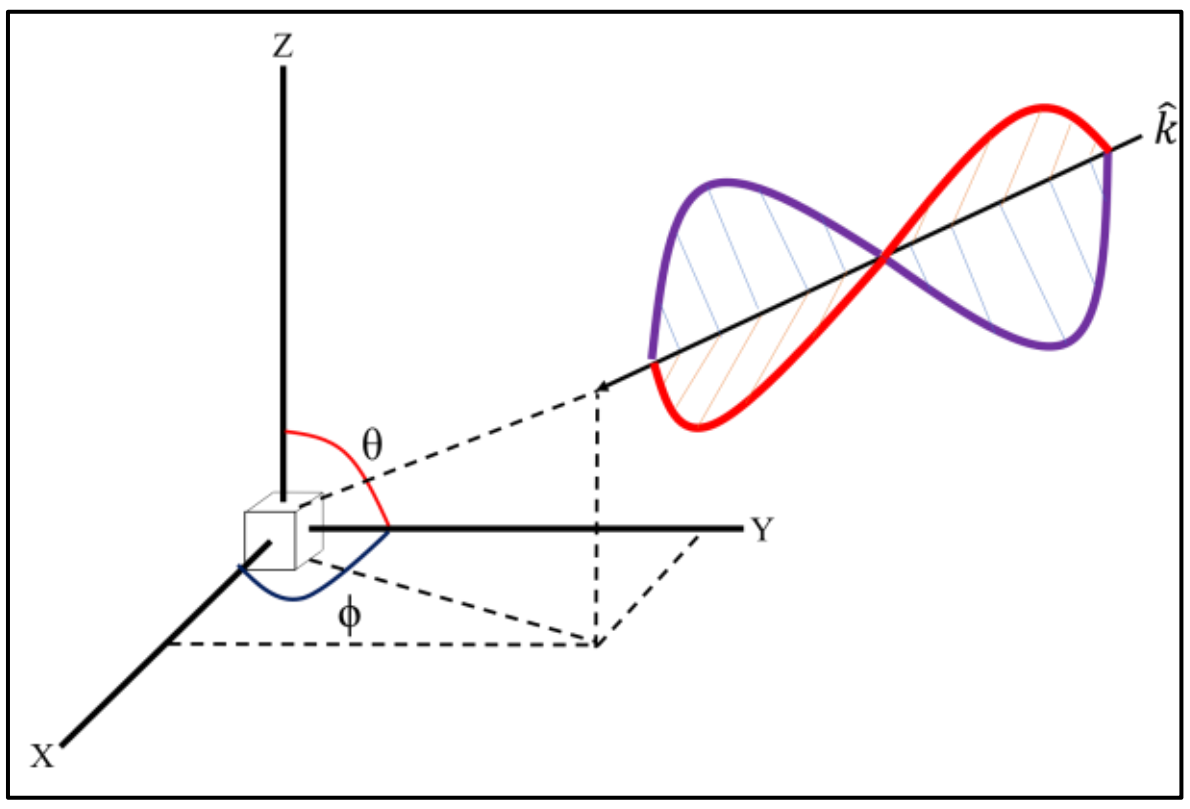

**Figure 1.** Tri-Monopole Antenna Configuration

The incident electromagnetic wave is assumed to originate from the far field, and the wavefront reaching the antenna is considered to be planar. Only one radio source is considered during each simulation, and multipath propagation and interference from other sources are not considered. Along with this, the polarization state of the incoming wave is assumed to remain constant during each observation interval. The simulations were performed using MATLAB, taking the effective length of the monopoles to be of unit length, using the steps shown in Figure 2.

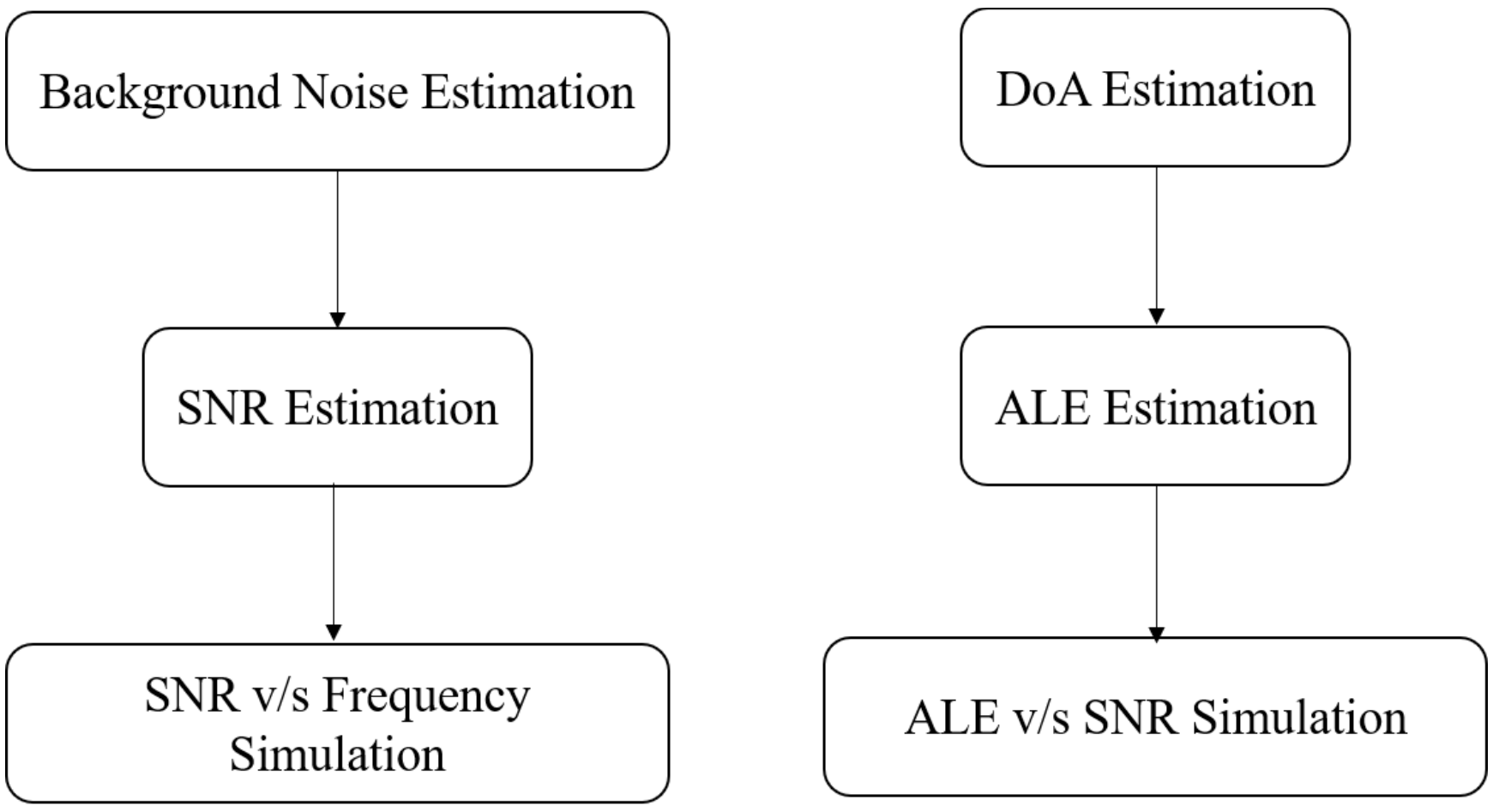

**Figure 2.** Methodology adopted for Simulations

## 4. Results & Discussion

The SNR vs Frequency plot shows how the quality of the received signal changes across the operating frequency range (300 kHz to 30 MHz) due to variations in antenna characteristics and background noise. It enables us to identify the frequencies at which the antenna receives signals with higher or lower SNR, indicating its effective operating range. For the background noise, the Galactic background noise is frequency dependent, and its contribution to the total noise changes across the spectrum. The QTN and receiver noise temperature are assumed to be constant, $10^4$ K and 300 K, respectively.

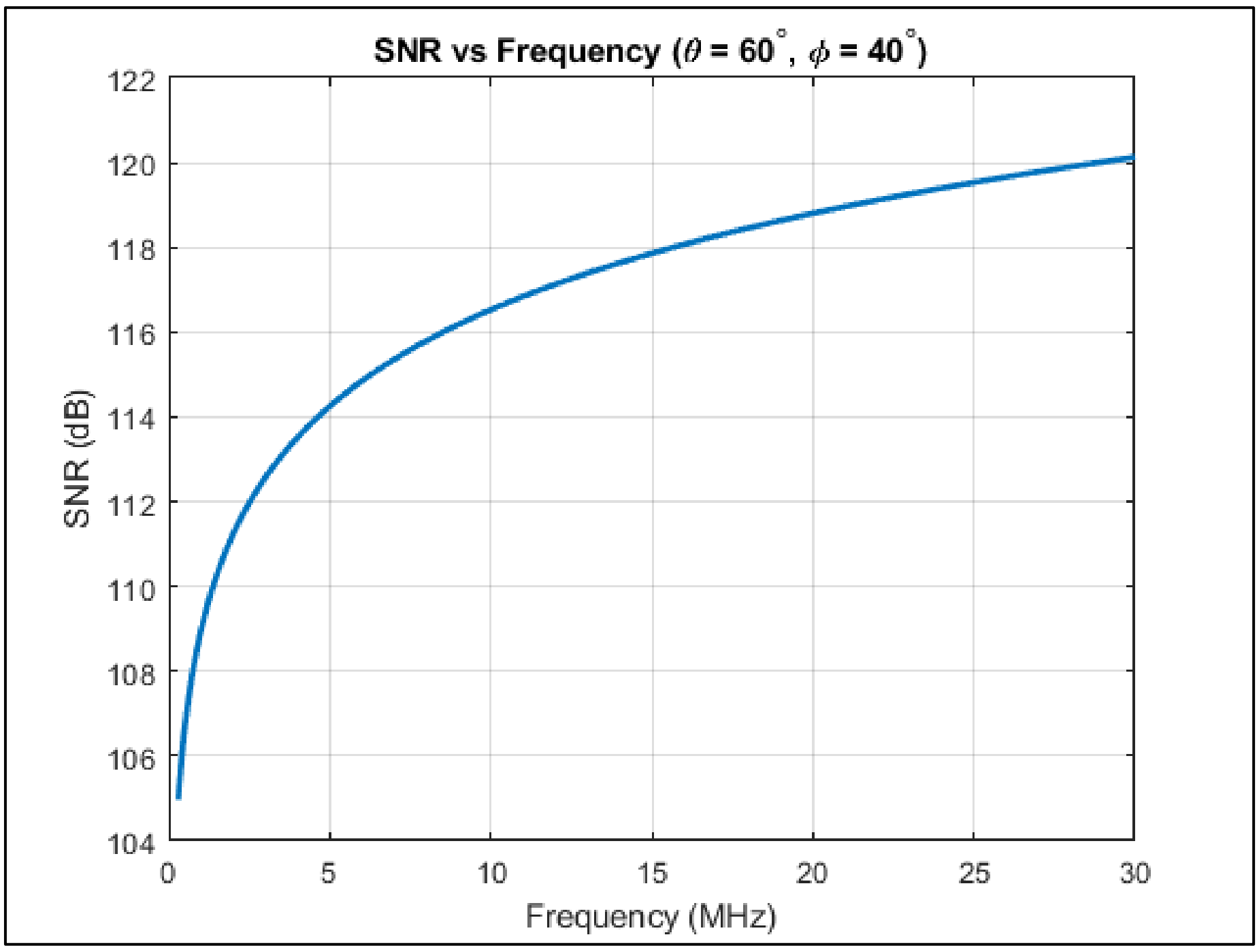

**Figure 3.** SNR vs Operating Frequency curve (for a specific value of θ and ϕ)

The SNR increases with frequency (as shown in Figure 3), that is, the received signal becomes progressively stronger relative to the background noise as the operating frequency increases. It implies that the antenna can distinguish the signal from the noise more effectively at higher frequencies within the frequency range considered [7]. The lower SNR observed at low frequencies is attributed to the electrically short antenna length relative to the wavelength, which reduces the antenna's effective length and received signal power, together with higher external noise (e.g., galactic, atmospheric, and man-made noise) and increased low-frequency receiver (1/f) noise. Thus, this suggests that the tri-monopole antenna is expected to provide better DoA estimation performance at higher frequencies within the analyzed frequency range.

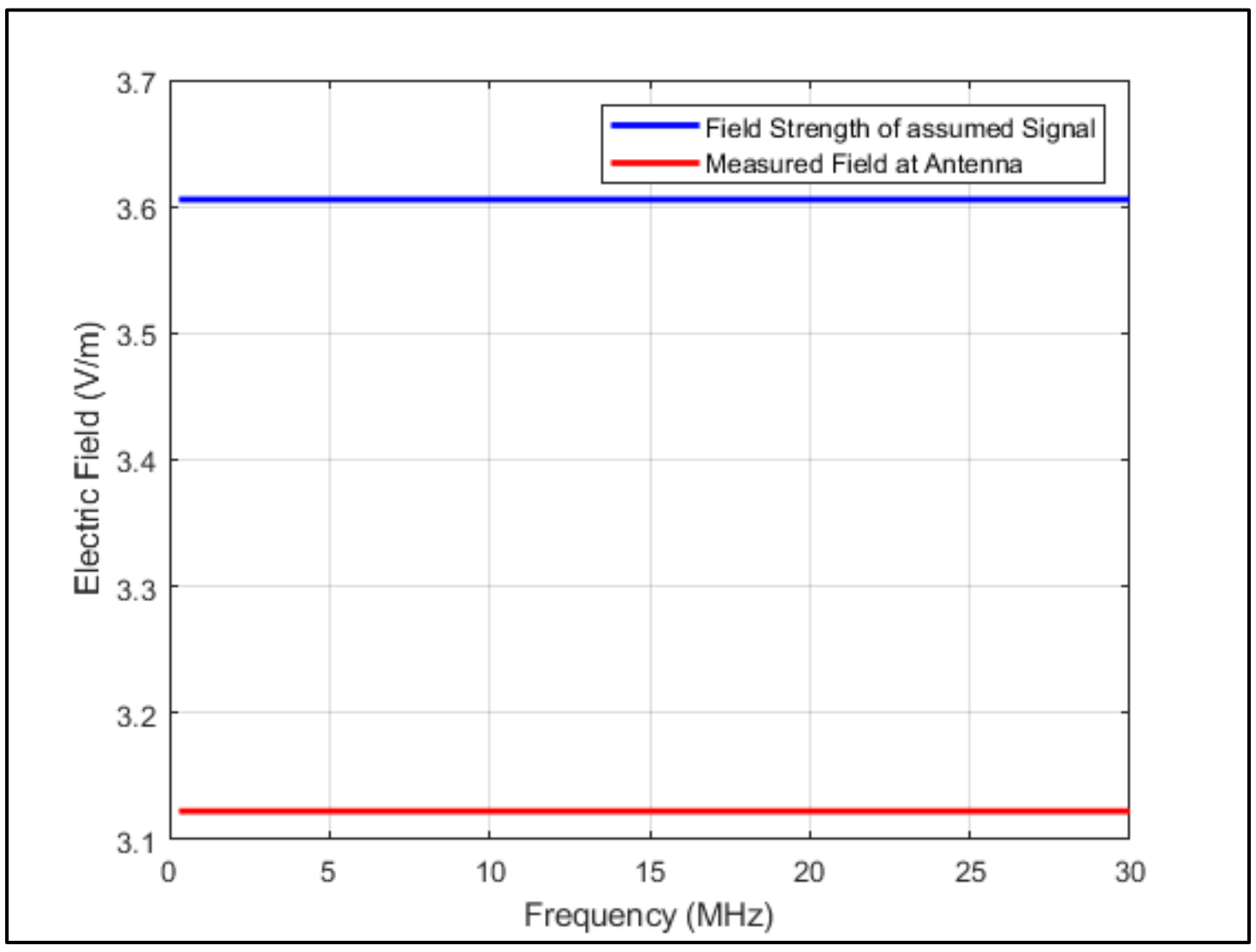

**Figure 4.** Field Strength of assumed signal and Measured field at one monopole

In the above Figure 4, the received antenna voltage is obtained by superimposing the incident signal voltage with additive complex Gaussian noise representing the combined effects of galactic, quasi-thermal, and receiver noise. The resulting noisy voltage is then used to determine the electric field measured at the antenna and for subsequent goniopolarimetric processing. The amplitudes of the orthogonal electric field components (A1 and A2) are assumed to be constant, representing a source with uniform field strength over the simulated frequency range. Consequently, the incident electric field remains constant with frequency, and the observed SNR variation arises solely from the frequency-dependent noise sources.

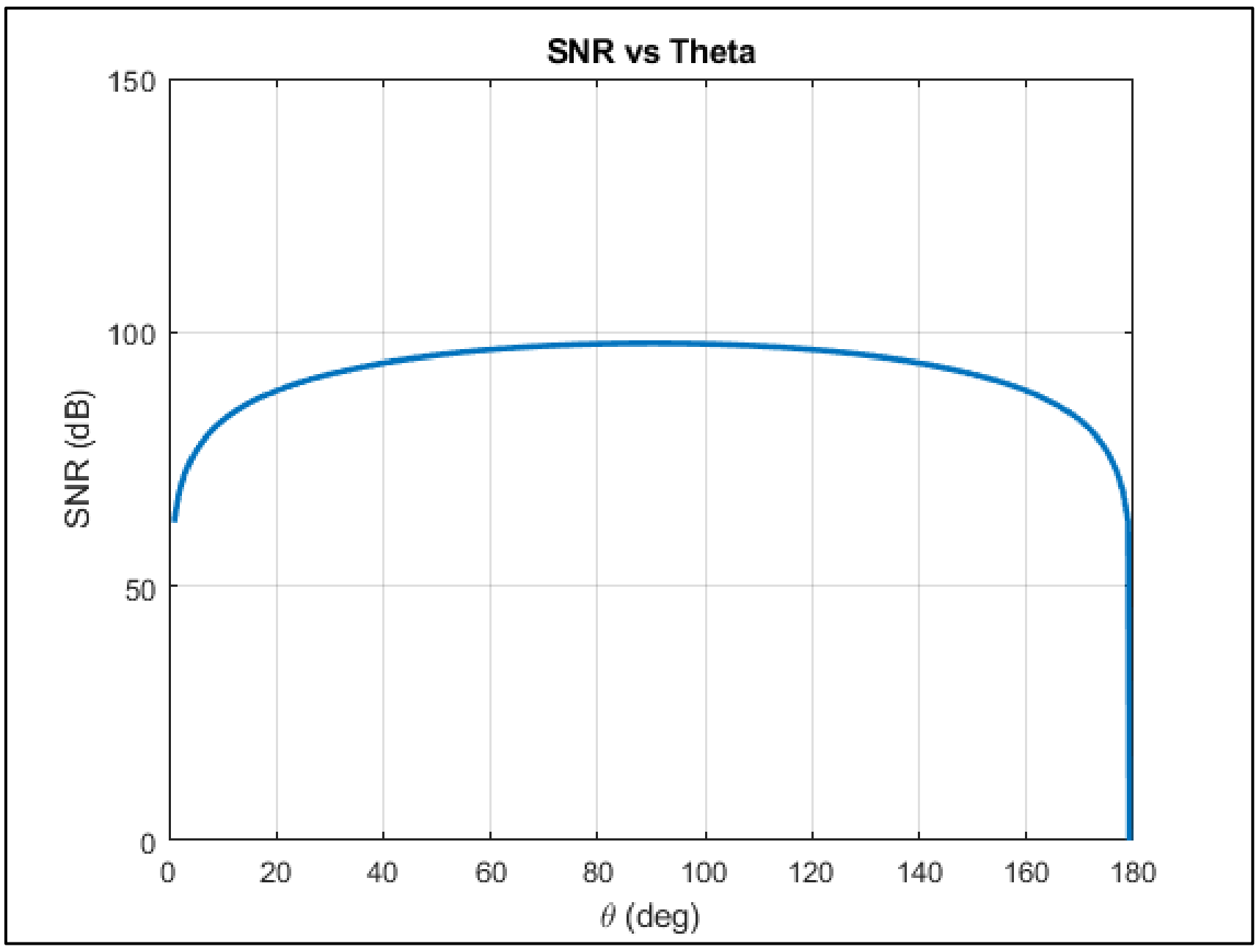

**Figure 5.** SNR variation at theta for Monopole (at 1 MHz)

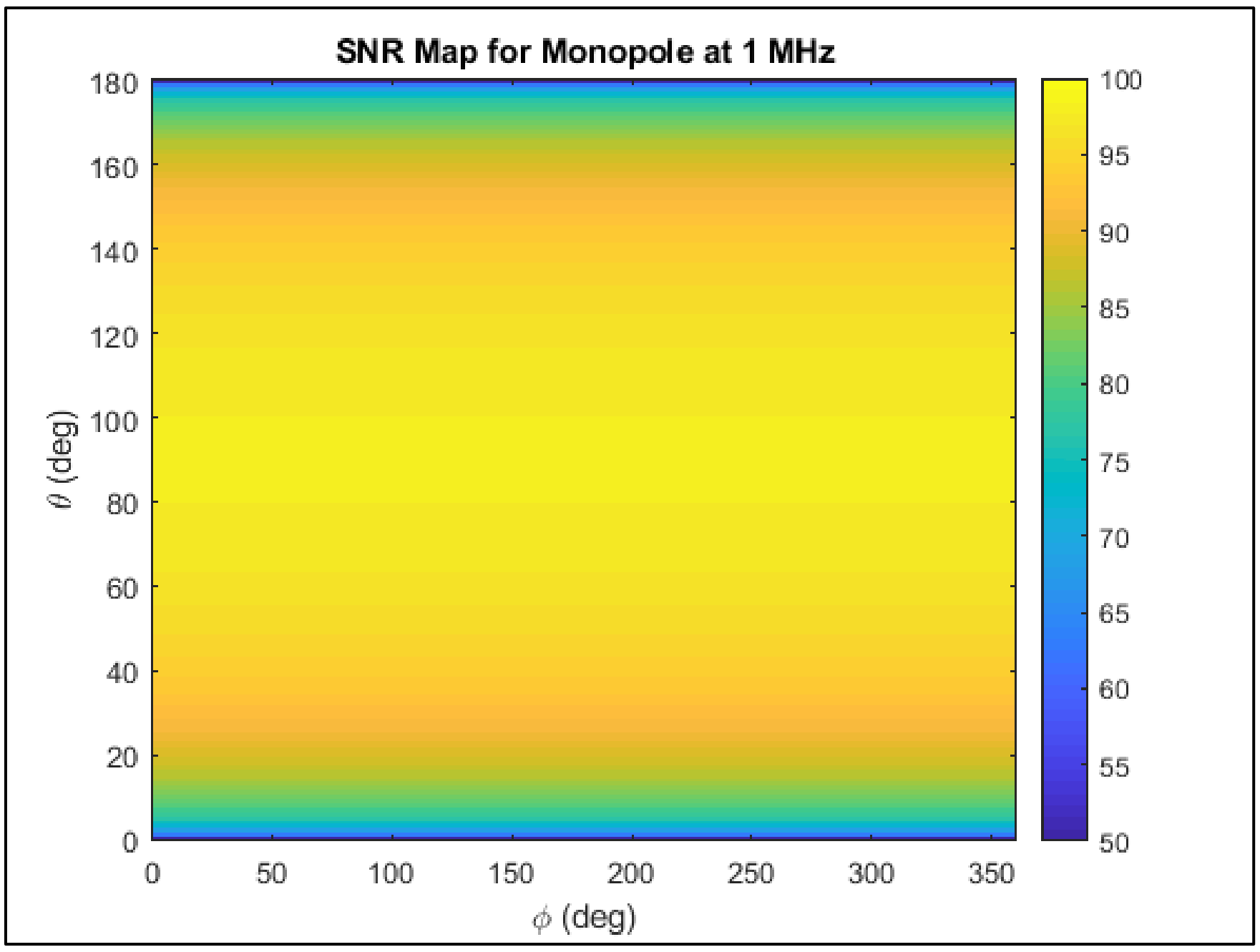

**Figure 6.** SNR variation for the complete Spatial Range for Monopole (at 1 MHz)

The SNR distribution over all azimuth (ϕ) and elevation (θ) angles at 1 MHz for monopole (as shown in Figures 5 and 6) and tri-monopole (as shown in Figures 7 and 8), provides a two-dimensional representation of the directional response of the antenna under background noise conditions. By evaluating the SNR for every possible arrival direction, the graph illustrates how effectively the antenna receives signals from different regions of space. Regions with high SNR indicate directions where the antenna is more sensitive, while regions with lower SNR correspond to reduced sensitivity. This helps characterize the angular reception pattern of the tri-monopole.

The angular SNR distribution provides insight into the expected localization performance. Directions with higher SNR are expected to yield lower ALE, whereas lower SNR regions may result in increased estimation uncertainty. Also, a more uniform SNR distribution indicates that the antenna can estimate the DoA with comparable accuracy over a wide range of incident angles. Any directional dependence observed in the graph reflects the inherent radiation and reception characteristics of the antenna geometry.

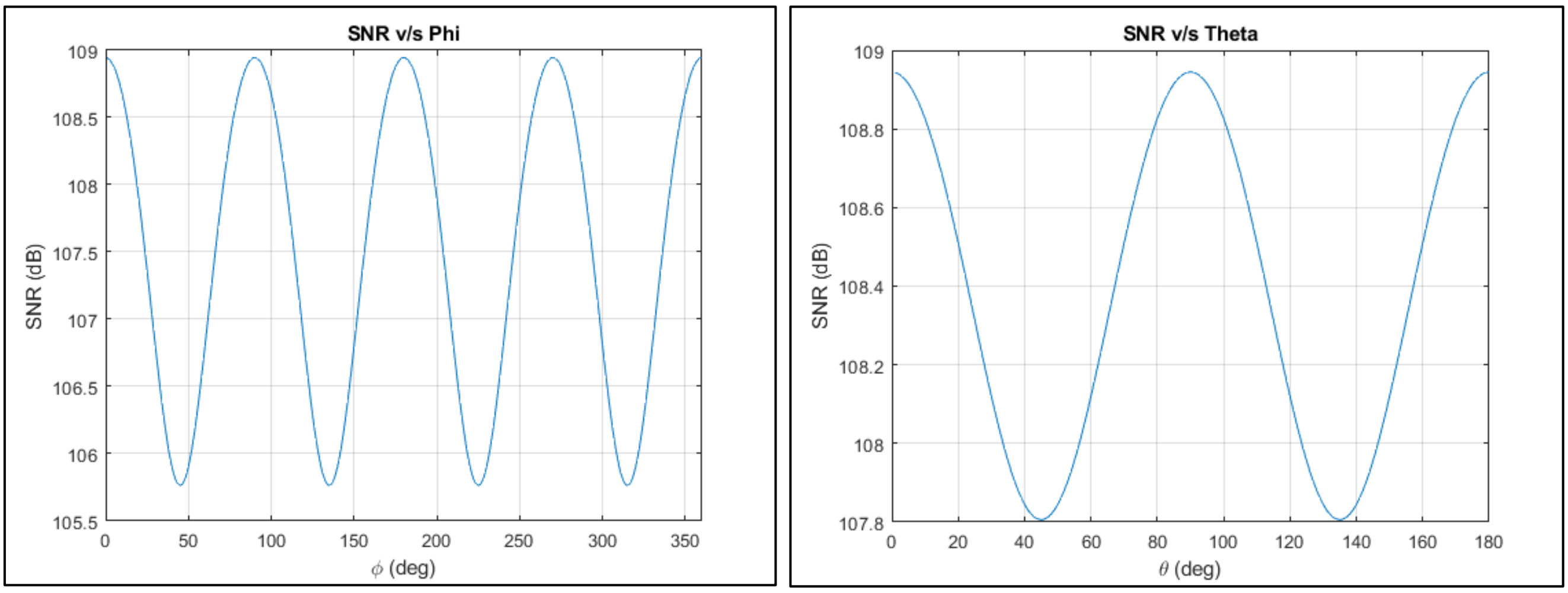

**Figure 7.** SNR variation at phi and theta for Tri-Monopole (at 1 MHz)

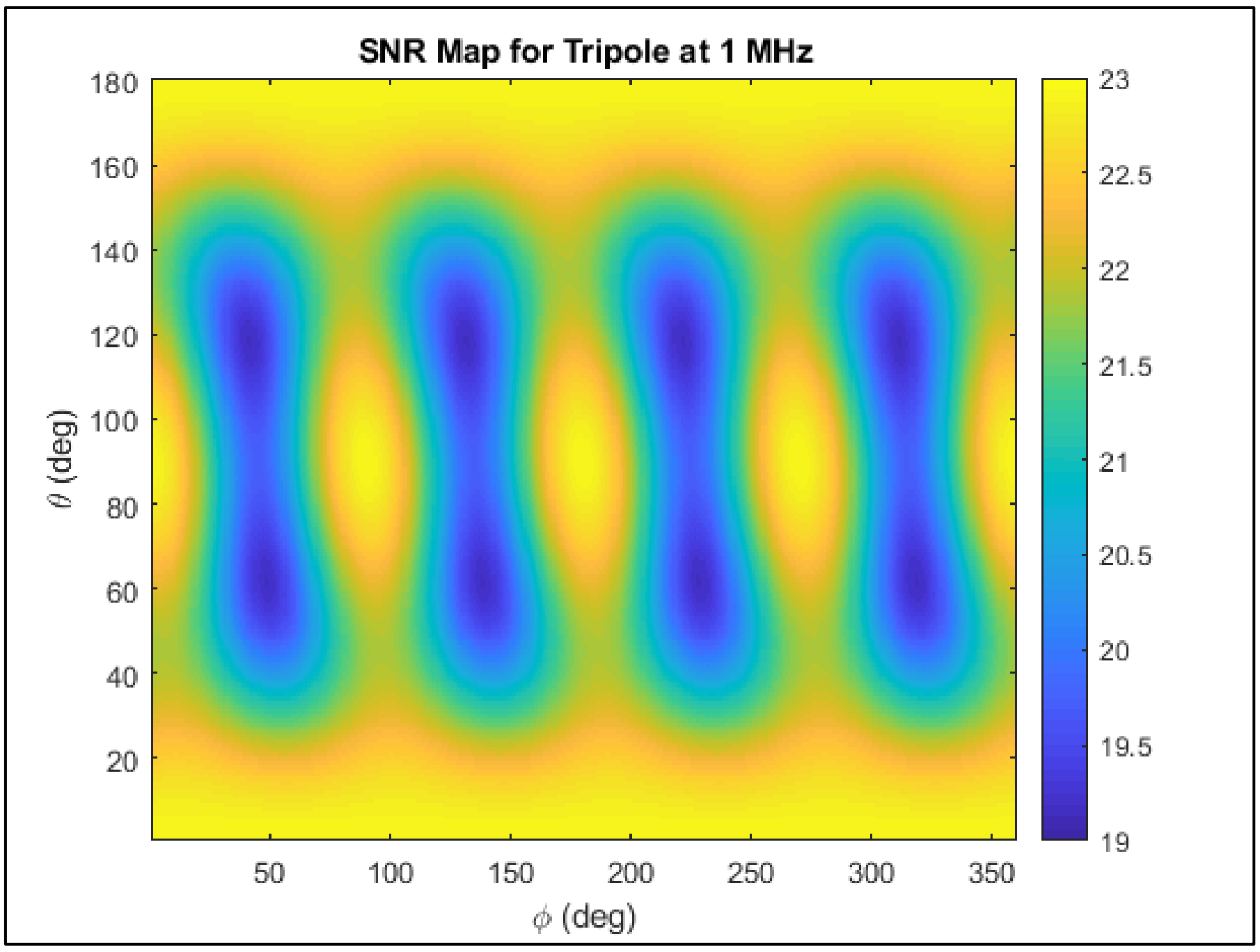

**Figure 8.** SNR variation for the complete Spatial Range for Tri-Monopole (at 1 MHz)

The graph also helps us to determine the minimum SNR required to achieve a desired localization accuracy, that is, ~ 1°. This information is valuable for selecting antenna parameters, receiver specifications, and operating conditions for practical goniopolarimetric systems.

At low SNR values, the noise significantly affects the received voltages, leading to larger errors in the estimated propagation vector and consequently higher ALE. As the SNR increases, the influence of noise is reduced, enabling more accurate estimation of the covariance matrix and the DoA. Therefore, the observed inverse relationship between ALE and SNR indicates that higher signal quality results in improved localization accuracy. This demonstrates the suitability of the proposed goniopolarimetric system for applications requiring accurate localization of low-frequency radio sources.

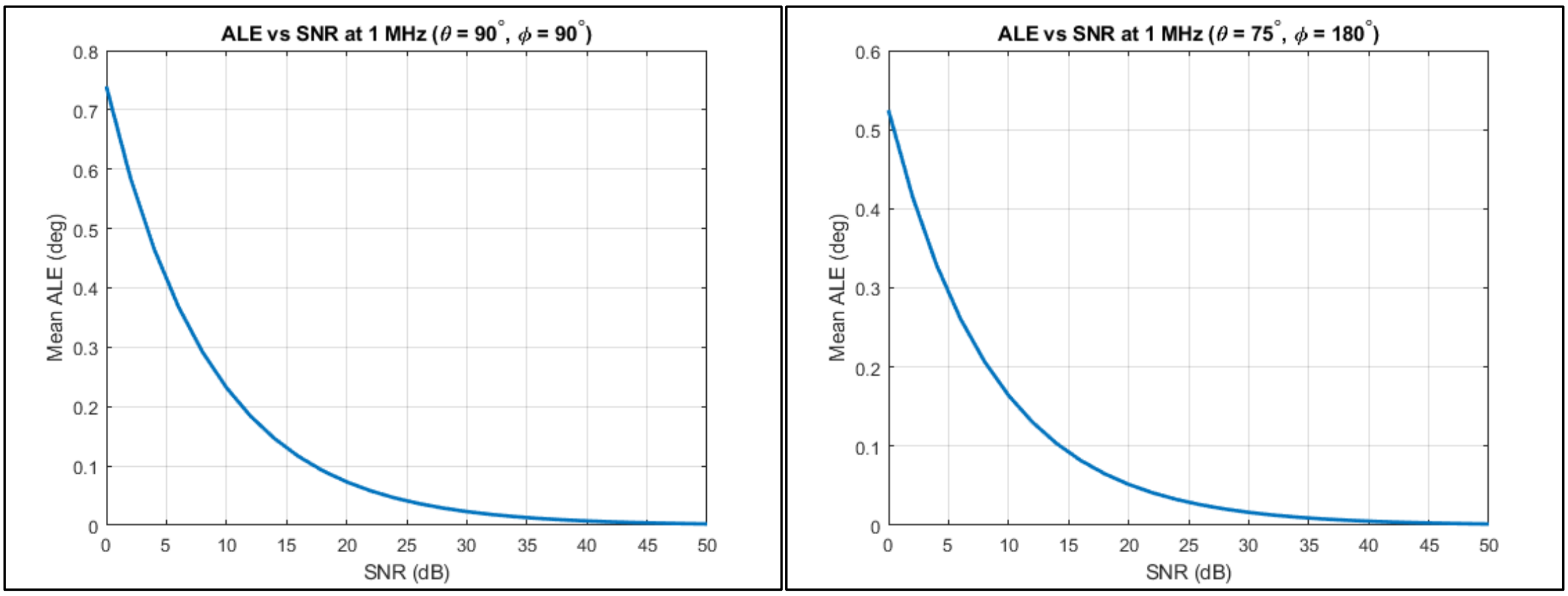

**Figure 7.** ALE with SNR for a specific value of θ and ϕ (at 1 MHz, ALE < 1°))

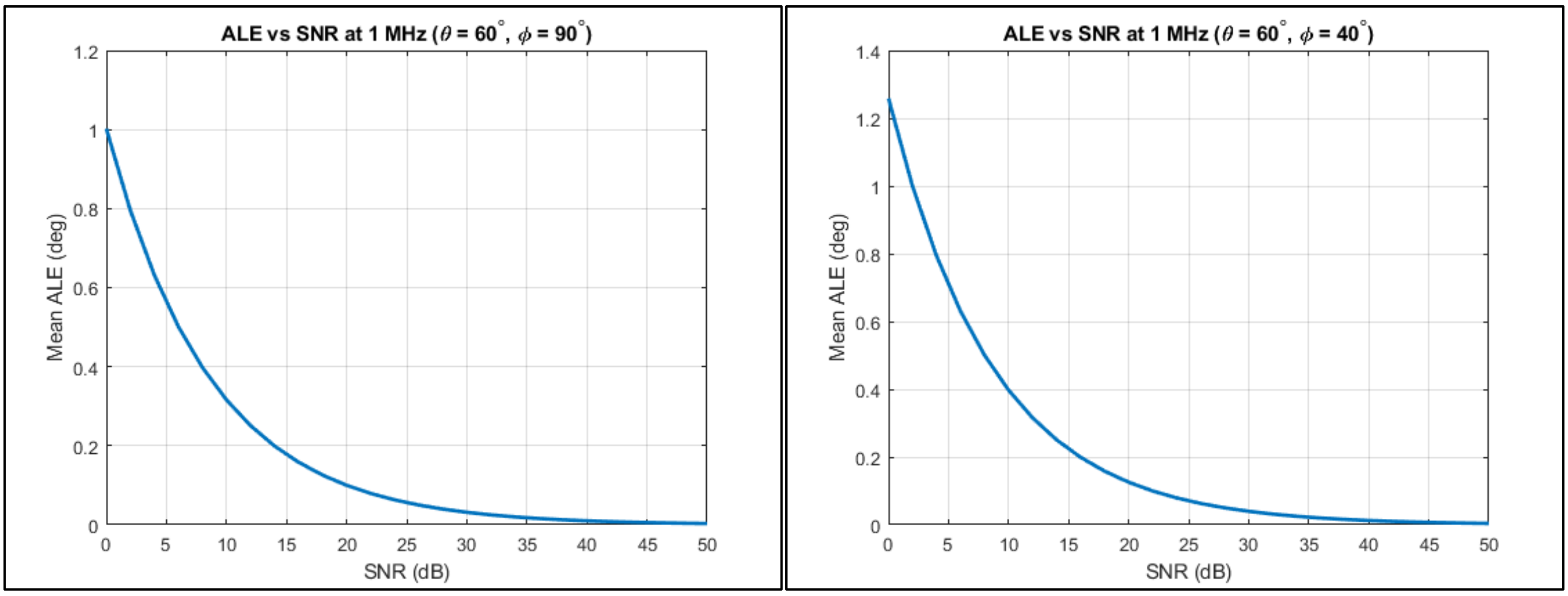

**Figure 8.** ALE with SNR for a specific value of θ and ϕ (at 1 MHz)

The ALE distribution over all azimuth (ϕ) and elevation (θ) angles at 1 MHz can identify how accurately the antenna estimates the DoA for different incident directions. Regions with lower ALE indicate higher localization accuracy, whereas regions with higher ALE correspond to directions where the estimation is less precise. Also, a symmetric distribution of ALE would indicate that the antenna configuration provides balanced directional performance, while localized increases in ALE reveal directions where the antenna is relatively less sensitive. The graph highlights the angular regions that provide the smallest localization errors. These regions represent the most favourable arrival directions for accurate goniopolarimetric measurements.

Although the ALE decreases with increasing SNR for all directions, its absolute value also depends on the direction of arrival relative to the tri-monopole antenna pattern. Although sub-degree ALE is achieved at 0 dB SNR for favourable arrival directions due to the robust directional response of the tri-monopole antenna, the simulation results indicate that an SNR greater than approximately 5 dB is required to consistently achieve optimum localization accuracy (ALE < 1°) across all θ and ϕ directions.

## 5. Conclusion

The simulation results in this study demonstrated the effectiveness of the proposed tri-monopole antenna for goniopolarimetric DoA estimation under background noise conditions. The SNR versus frequency analysis shows that the received signal becomes more distinguishable due to decreased galactic noise and that the antenna operates more effectively at higher frequencies within the simulated band. The SNR distribution over all azimuth and elevation angles confirms the directional reception characteristics of the antenna and identifies the angular regions that provide favourable signal quality for DoA estimation. The corresponding ALE distribution over θ and ϕ demonstrates that the localization accuracy varies with the arrival direction of the incident wave, providing a comprehensive assessment of the antenna's directional localization capability. The ALE versus SNR analysis establishes an inverse relationship between signal quality and localization error, indicates that improved signal quality leads to more accurate estimation of the wave propagation vector and, consequently, more precise DoA estimation.

These findings demonstrate that the proposed tri-monopole antenna is capable of achieving reliable goniopolarimetric DoA estimation under galactic noise-limited conditions and provide a strong foundation for future investigations incorporating practical effects such as receiver noise, quasi-thermal noise, antenna calibration errors, and space plasma environments. Future work on how ALE will change with partially polarized EM waves will also be performed.

## 6. Acknowledgements

The Space Weather RAdio (SWARA) Polarimeter is being developed at the Indian Institute of Technology (IIT), Kanpur in collaboration with the Physical Research Laboratory (PRL). This work was carried out as part of the collaborative project between IIT Kanpur and PRL.